\documentstyle[preprint,aps]{revtex}

\begin{document}

\preprint{\parbox{2in}{
\begin{flushright} UMHEP-421 \\ hep-ph/9509240
\end{flushright}
}}

\title{ Global anomalies and the gauge-boson equivalence theorem }
\author{ John F. Donoghue\thanks{E-mail: donoghue@phast.umass.edu.}
 and Jusak Tandean\thanks{E-mail: jtandean@phast.umass.edu.} }
\address{Department of Physics and Astronomy \\
University of Massachusetts \\ Amherst, Massachusetts 01003, USA}

\maketitle

\begin{abstract}
We discuss the various resolutions which have been suggested in
the literature for the way that the equivalence theorem can be
satisfied in theories with global anomalies.
We provide a model-independent proof for the resolution originally
suggested in the context of a toy model by Kilgore.
\end{abstract}



The equivalence theorem for gauge bosons~\cite{et} states that
high-energy matrix elements  involving longitudinal gauge bosons are
equal to  similar matrix elements  involving the \mbox{(pseudo-)}
scalar Goldstone bosons of the theory, up to corrections of order
$M_W/\sqrt{s}$.
In a previous paper~\cite{dt} we raised a concern about the equivalence
theorem in theories with global anomalies, in which the anomalies are
required to cancel in gauge currents but not in the coupling of the
Goldstone bosons.
{}From this we argued that, for example, the $\gamma^*\gamma Z_{\rm L}$
coupling vanished, but that the $\gamma^*\gamma z$ coupling did not,
where $z$ is the  Goldstone boson absorbed by the $Z$ boson.
(These could in principle be compared in the reactions
$\; e^+ e^- \rightarrow \gamma Z_{\rm L} \;$ and
$\; e^+ e^- \rightarrow \gamma z \;$  at a high energy.)
Two papers written in response proposed diametrically opposite
solutions.
Pal~\cite{pal} has argued that the equivalence theorem is restored in
this example by the vanishing of the $\gamma^*\gamma z$ coupling, so
that the anomaly generates neither process.
On the other hand, Kilgore~\cite{kilgore} used a quark model for chiral
symmetry breaking to argue that at the relevant energies both the
$\gamma^*\gamma Z_{\rm L}$ and $\gamma^*\gamma z$ couplings were
nonvanishing (and equivalent).
In this paper we revisit the issue and provide a model-independent
derivation which supports Kilgore's result.
This resolution has the interesting consequence of allowing unexpected
anomaly-generated vertices of  gauge bosons.

To start out, let us define a toy model where there are anomalous
couplings of
the Goldstone bosons, but the gauge currents have no anomalies.
This requires that the Goldstone bosons be formed from only a subset
of the
fermions in the theory, and that other fermions exist for cancelling
anomalies in the gauge currents.
The analogy is with the low-energy  sector of the Standard Model, where
both  quarks and leptons are coupled to the gauge currents, but only
quarks form the pseudo-Goldstone bosons (pions).
For the toy model here we consider a doublet of techniquarks and
a doublet of
leptons, all of which are involved in electroweak interactions.
The techniquarks are also involved in technicolor interaction,
which is taken
to be QCD-like, i.e. confining and with dynamical symmetry breaking.
If the quantum numbers of the fermions are chosen to be the same as
those of the usual single family of quarks and leptons, then the
Goldstone bosons will
have the same anomalous couplings to $\gamma$, $W$, $Z$ as does the
pion in QCD.
However, the leptons cancel all anomalies in the gauge currents.
All the fermions are strictly massless, as to introduce a mass would
require a new set of interactions and would seriously complicate
the analysis.

In such a theory there is a window of energies where it is useful
to apply the equivalence theorem.
At low energies, the theorem is not useful because the corrections,
of order $M_W/\sqrt{s}$, are too large, so this fact requires
$\; s\gg M^2_W .\;$
At the high-energy end, at energies much larger than the
technicolor scale,
the dynamics becomes that of the techniquarks rather than of the bound
technihadrons, so that while the equivalence theorem may still be
applied, we
lack the tools to predict the Goldstone-boson amplitudes.
(The analogy would be trying to predict pionic amplitudes in QCD at
energies $\; E\gg 1\,{\rm GeV}.)\;$
For intermediate energies,
$\; (250\,{\rm Gev})^2 \;^{_{\textstyle <}}_{^{\textstyle\sim}}\; s
   \;^{_{\textstyle <}}_{^{\textstyle\sim}}\; M^2_{\rho_T} ,\;$
where  $\; M_{\rho_T}\simeq 2\,{\rm TeV} \;$ is the technirho mass,
one may use the techniques of chiral perturbation theory to easily
predict Goldstone-boson
amplitudes in this type of technicolor model.
We therefore restrict our comments to energies high
enough that the  ${\cal O}(M_W/\sqrt{s})$ corrections can be
neglected, but low enough that we can use effective Lagrangians to
describe the Goldstone-boson couplings.

In this toy variant of technicolor theories, we expect that the
neutral  Goldstone boson, the technipion $\pi_T^0$, would have
a coupling to two photons, $\; \pi_T^0 \rightarrow \gamma\gamma ,\;$
similar to that which occurs in QCD  and  given by
\begin{eqnarray}
{\cal M}_{\pi^0_T \rightarrow \gamma\gamma} \;=\;
 -{e^2 N_{\rm TC} \over 12\pi^2 F_{\pi_T}} \,
  \epsilon^{\mu\nu\alpha\beta} \,
 q_\mu q'_\nu \,\epsilon_\alpha^\ast (q) \epsilon_\beta^\ast (q') \;,
\label{pgg}
\end{eqnarray}
where $\;F_{\pi_T}\simeq 250\,{\rm GeV}\;$ is the technipion decay
constant  and  $N_{\rm TC}$ is the number of technicolors in an
SU($N_{\rm TC}$) theory.
For this toy model $N_{\rm TC}=3 .\;$
There are also related  $\; \pi_T^0 \rightarrow \gamma Z,\, ZZ \;$
couplings.
One cannot apply the equivalence theorem in decay processes such as
$\; \pi_T^0 \rightarrow \gamma\gamma \;$
(note that $\; Z_{\rm L} \rightarrow \gamma\gamma \;$  is forbidden
by Yang's theorem \cite{yang})
because the energy involved is too low.
However, one can use this vertex within a high-energy diagram such
as
$\; e^+ e^- \rightarrow \gamma^*,\, Z^* \rightarrow \gamma \pi^0_T \;$
or
$\; \nu \bar{\nu} \rightarrow Z^* \rightarrow \gamma \pi^0_T .\;$
These reactions involve an off-shell photon or $Z$.
There will then be
higher-order corrections to the amplitude~(\ref{pgg}) which depend on
the  $\;q^2 \equiv s \;$ of this off-shell particle, and we can
represent the change in the amplitude by a form factor $F(s)$, with
$\;F(0) = 1 .\;$
In the QCD analogy, this form factor has been measured in the
reactions $\; \gamma^*\gamma \rightarrow \pi^0,\,\eta \;$ studied
at $e^+ e^-$ colliders~\cite{expt}, and, not  surprisingly, it varies
with s, with a scale around $1\,{\rm GeV}$.
(More accurately, it behaves roughly as  $1/(1-s/M^2_V)$  where
$\;M_V = M_\rho,\;$ with $M_\rho$ being the $\rho$-meson mass.)
In the technicolor case, one would expect that the form factor would
vary in a
similar way, with the relevant scale being the mass of the technirho.
This leads to an enhancement of the rate at the energies which we are
considering since $\;F(s) > 1 \;$  for $s$ timelike.

The anomalous gauge couplings $\gamma^*\gamma Z_{\rm L}$ and
$Z^*\gamma Z_{\rm L}$
do not occur at tree level, and naively one would expect them not to
occur at all since all anomalies cancel in the gauge currents.
However, since the $\pi^0_T$ is the Goldstone boson associated with
the $Z$,  this would contradict the equivalence theorem.
In order to resolve the apparent contradiction,
either the $\gamma^*\gamma \pi^0_T$ vertex has to be shown to vanish
or the $\gamma^*\gamma Z_{\rm L}$ vertex should be
nonvanishing.\footnote{In the rest of
the paper we will deal with  only the $\gamma^*\gamma Z_{\rm L}$ and
$\gamma^*\gamma \pi^0_T$ vertices because the same discussion could
be given  for the $Z^*\gamma Z_{\rm L}$ and
$Z^*\gamma \pi^0_T$ vertices.}

The paper by Pal~\cite{pal} claims that the first of the above
possibilities is correct, i.e. that the $\gamma^*\gamma\pi^0_T$
vertex vanishes for a virtual photon with $\;q^2 \neq 0.\;$
The key step in his argument is a representation of the triangle
diagram for a free techniquark of mass $m$ with one photon off-shell,
at $\;p_{\pi_T}^2 = 0,\;$  where $p_{\pi_T}$ is the momentum of the
technipion,
\begin{eqnarray}
{\cal M}_{\pi^0_T \rightarrow \gamma^*\gamma} \;=\;
-{e^2 \over 4\pi^2 F_{\pi_T}} \, \epsilon^{\mu\nu\alpha\beta} \,
 q_\mu q'_\nu \,\epsilon_\alpha^\ast (q) \epsilon_\beta^\ast (q')
\Bigg( 1 + 2 q^2 \int_0^1 {\rm d}z \int_0^{1-z} {\rm d}z' \,
      {z z' \over m^2 - q^2 zz'} \Bigg) \;.  \label{pal}
\end{eqnarray}
The idea is that at $\;m \neq 0\;$, $\;q^2 = 0\;$  (as in the case of
QCD with real photons) the factor of unity reproduces the usual
$\; \pi_T^0 \rightarrow \gamma\gamma \;$ result, but that at
$\;m = 0,\;$  $\;q^2 \neq 0\;$  the amplitude vanishes.
However, it is important to note that this result is appropriate only
for free, unconfined fermions.
The integral has a cut starting at $\;q^2 = 4m^2 ,\;$  which
corresponds to the productions of techniquark pairs.
For QCD-like theories, this representation is contradicted by the
known measurements of the form factor of the anomaly mentioned above.
In QCD, the representation of Eq.~(\ref{pal}) predicts that at
$\;q^2 > m^2 = m_q^2 \;$  (with $\;m_q \sim 10 \,\rm MeV $)  the
anomaly
amplitude should fall off and quickly vanish as $q^2$ increases,
i.e. that the scale of the $q^2$ variation of the form factor is the
light-quark mass $m_q$.
This is experimentally not correct, and the theoretical reason for
this is easily understood.
In a confining theory such as QCD, there is no long-range
propagation of quarks, so there is little variation of the amplitude
for  $\;q^2 \simeq m_q^2\;$  when $m_q$ is small compared to the
energy scale of the theory.
The relevant energy scale is that of the boundstates, which in this
channel are the vector mesons.
The representation~(\ref{pal}) is not appropriate for a confining
theory, so that it does not apply to our technicolor theory.

This conclusion can be demonstrated more explicitly by a
consideration of the effective Lagrangian for the anomaly.
This yields an expansion of the effective action in powers of $q^2$
and $m_q$, and can be used to show that there is no
problem with the $\;m_q \rightarrow 0 \;$  limit.
An effective Lagrangian must contain all of the
fields that are light at the energy scale under discussion.
In a free quark theory  this would include the quark fields, but
in a confining theory the quarks are integrated out and only the pion
fields enter the effective Lagrangians.
At lowest order the effect of the anomaly is contained in the
gauged Wess-Zumino anomaly Lagrangian~\cite{wz,witten,wzw,wzw0}.
This will be written more fully later in this paper, but the relevant
terms for the present discussion are
\begin{eqnarray}
{\cal L}_{\rm WZ} \;=\;
- {{\rm i} e^2 \over 8\pi^2} \, \epsilon^{\mu\nu\alpha\beta} \,
 \partial_\mu A_\nu A_\alpha \, {\rm Tr}
\Big[ Q^2 \big( \partial_\beta U U^\dagger
               + U^\dagger \partial_\beta U \big) +
     \mbox{$1\over2$} \big( Q\partial_\beta U Q U^\dagger
                - Q\partial_\beta U^\dagger Q U \big)
\Big]   \;,
\end{eqnarray}
where $\;U=\exp( {\rm i} \tau^a\pi^a/F_\pi ),\;$ with $F_\pi$ being
the pion decay constant and  $\tau^a$ ($ a=1,2,3 $) being Pauli
matrices,  and  $Q$ is the quark-charge matrix.
The coefficient is fixed and independent of the quark  mass.
At next order there are many different Lagrangians which modify
$\; \pi^0 \rightarrow \gamma\gamma .\;$
These are not technically anomalous~\cite{wzw2}, but they do involve
the $\epsilon_{\mu\nu\alpha\beta}$ tensor.
Examples are
\begin{eqnarray}
{\cal L}^{(6)}_\epsilon  &=&
{{\rm i} e^2 \over \Lambda_\chi^2} \, \epsilon^{\mu\nu\alpha\beta} \,
\partial^\lambda F_{\lambda\mu} F_{\alpha\beta} \,
\Big[
c_1 \, {\rm Tr} \big( Q^2 U^\dagger \partial_\nu U
                     - Q^2 U \partial_\nu U^\dagger \big)
+ c_2 \, {\rm Tr} \big( Q U^\dagger Q\partial_\nu U
                       - Q U Q\partial_\nu U^\dagger \big)
\Big]
\nonumber \\ &&
+\;
{{\rm i} e^2 \over \Lambda_\chi^2} \, \epsilon^{\mu\nu\alpha\beta} \,
F_{\mu\nu} F_{\alpha\beta} \,
\Big[ c_3 \, {\rm Tr} Q^2 \, {\rm Tr} \big( m_q U - m_q U^\dagger \big)
     + c_4 \, {\rm Tr} \big( Q^2 m_q U^\dagger - Q^2 m_q U \big)
\Big]  \;,
\end{eqnarray}
where $\; F_{\mu\nu} = \partial_\mu A_\nu - \partial_\nu A_\mu .\;$
The  coefficients    are also independent of $q^2$ and $m_q$,
as all of the energy and mass dependence  is accounted for explicitly
in the construction of  ${\cal L}^{(6)}_\epsilon$.
The energy scale $\Lambda_\chi$ in these coefficients is of order
the QCD scale, $\; \Lambda_\chi \sim M_\rho .\;$
The matrix elements for off-shell photons can be read off of these
Lagrangians.
Since the coefficients are independent of $m_q$ and there are no light
particles in  the theory which need to be accounted for (aside from
the pions),
the amplitude is smooth in the limit $\;m_q \rightarrow 0.\;$

The opposite resolution, that the $\gamma^*\gamma Z_{\rm L}$ coupling is
nonzero, has been suggested by Kilgore~\cite{kilgore}.
He employs a quark model in which, by a transformation on the original
techniquark fields, one can generate constituent techniquarks with
a large mass $M_Q$ coupled to technipions.
The anomalous $\gamma^*\gamma \pi^0_T$  coupling is determined by the
techniquarks.
However, in the $\gamma^*\gamma Z_{\rm L}$ coupling Kilgore found
that the massive techniquarks do not contribute at low energies,
$\; M_Z^2 \ll s \ll M_Q^2 ,\;$
while the leptons still do, yielding a net
$\gamma^*\gamma Z_{\rm L}$ coupling whose magnitude is the same as
the   $\gamma^*\gamma \pi^0_T$ vertex.

We would like to demonstrate that this resolution is correct, in a
model-independent way.
This can be accomplished by showing that, after the techniquarks have
been integrated out, the  $\gamma^*\gamma Z_{\rm L}$ vertex comes from
either both the lepton sector and the anomalous effective Lagrangian
or  the lepton
sector only, depending on the renormalization prescription one uses.
In both cases the  $\gamma^*\gamma\pi^0_T$ vertex is unchanged, and is
equivalent to the  $\gamma^*\gamma Z_{\rm L}$ vertex.

The full effective Lagrangian describing the effect of  anomalies is
determined by anomalous Ward identities and was first given in a
power-series representation by Wess and Zumino~\cite{wz}.
Subsequently, Witten~\cite{witten} showed (in the context of QCD) how
the pionic portion could be represented as an integral over
a five-dimensional space, and also gave the four-dimensional anomalous
coupling of the electromagnetic field with pions.
Several authors~\cite{wzw,wzw0} have corrected an error in Witten's
form and  have given the full anomalous Lagrangian describing
the couplings of the gauge fields.
The total result is very lengthy, but the portion containing
one pseudoscalar meson and two electroweak gauge bosons is,
in the context of technicolor,
\begin{eqnarray}
{\cal L}_{\rm WZ}  \;=\;
- {{\rm i} N_{\rm TC} \over 48 \pi^2}\, \epsilon^{\mu\nu\alpha\beta} \,
{\rm Tr} \bigl[ &&
\bigl( \partial_\mu \ell_\nu \, \ell_\alpha
      + \ell_\mu \, \partial_\nu \ell_\alpha \bigr)
 \bigl( \partial_\beta U \, U^\dagger - iU r_\beta U^\dagger \bigr)
\nonumber \\ &&
+\;
\bigl( \partial_\mu r_\nu \, r_\alpha
      + r_\mu \, \partial_\nu r_\alpha \bigr)
 \bigl( U^\dagger \, \partial_\beta U
       + {\rm i} U^\dagger \ell_\beta U \bigr)
\nonumber \\ &&
-\; \partial_\mu \ell_\nu \, \partial_\alpha U \, r_\beta U^\dagger
+ \partial_\mu r_\nu \, \partial_\alpha U^\dagger \, \ell_\beta U \;
\bigr]  \;,
\end{eqnarray}
where
\begin{eqnarray}
\ell_\mu \;=\;
e Q A_\mu  + {e \over c_{\rm w} s_{\rm w}} (T_3 - s_{\rm w}^2 Q) Z_\mu
+ {e\over 2s_{\rm w}} (\tau_1 W_{1\mu} + \tau_2 W_{2\mu})
\;,&&   \;\;\;
r_\mu \;=\; eQA_\mu - {e s_{\rm w}\over c_{\rm w}}\, QZ_\mu  \;,
\end{eqnarray}
with  $\; s_{\rm w} = \sin{\theta_{\rm W}} \;$
($\theta_{\rm W}$ is the Weinberg angle)
and   $\; c_{\rm w} = \sqrt{1-s_{\rm w}^2} ,\;$
and now $\; U=\exp( {\rm i} \tau^a\pi^a_T/F_{\pi_T} ) .\;$

The total expression for ${\cal L}_{\rm WZ}$ corresponds to an
anomalous effective  action whose variation under an infinitesimal
chiral gauge transformation  yields the so-called left-right (LR) form
of the (non-Abelian) anomaly~\cite{wzw,wzw0}.
In deriving the LR anomaly, one employs a renormalization
prescription in which the non-Abelian left- and right-handed currents
are treated symmetrically,
and so  both chiral currents have anomalous divergences.
${\cal L}_{\rm WZ}$ contains a $\gamma^*\gamma Z_{\rm L}$ coupling,
which is to be added to the contribution of the leptons  to the
$\gamma^*\gamma Z_{\rm L}$ vertex.
The coupling is given by
\begin{eqnarray}
{\cal M}_{Z_{\rm L} \rightarrow \gamma^*\gamma}^{\rm WZ} \;=\;
 -{{\rm i} e^3 \over 12\pi^2 c_{\rm w} s_{\rm w} M_Z} \,
\epsilon^{\mu\nu\alpha\beta} \,
 q_\mu q'_\nu \,\epsilon_\alpha^\ast (q) \epsilon_\beta^\ast (q')
\;+\; {\cal O}(M_Z/E_Z)  \;.
\label{mwz}
\end{eqnarray}
In this renormalization scheme, the left- and right-handed leptons
contribute seperately to the  $\gamma^*\gamma Z_{\rm L}$ vertex.
Each lepton triangle involves chiral currents at each vertex and
is required  to be symmetric under interchange of any  pair of
vertices~\cite{pok}.
One then gets
\begin{eqnarray}
{\cal M}_{Z_{\rm L} \rightarrow \gamma^*\gamma}^{\rm lepton} \;=\;
 -{{\rm i} e^3 \over 24\pi^2 c_{\rm w} s_{\rm w} M_Z} \,
\epsilon^{\mu\nu\alpha\beta} \,
 q_\mu q'_\nu \,\epsilon_\alpha^\ast (q) \epsilon_\beta^\ast (q')
\;+\; {\cal O}(M_Z/E_Z)  \;.
\label{ml}
\end{eqnarray}
In both~(\ref{mwz}) and~(\ref{ml}) we have used the fact that
$\; \epsilon^{Z_{\rm L}}_\mu(p) \simeq p_\mu/M_Z
   + {\cal O}(M_Z/E_Z) \;$
for   $\; E_Z \gg M_Z .\;$
With the replacement $\; M_Z = eF_{\pi_T}/(2c_{\rm w}s_{\rm w}) ,\;$
we see that at these energies the sum of the amplitudes in~(\ref{mwz})
and~(\ref{ml}) is equal  in magnitude to
${\cal M}_{\pi^0_T \rightarrow \gamma^*\gamma}$, given by the
right-hand side of Eq.~(\ref{pgg}).

The same result can be derived using the Bardeen form of the anomaly
\cite{bardeen,wzw}, and the way that this is accomplished
is interesting.
This scheme is equivalent to the regularization employed in
Kilgore's model.
In the Bardeen form, one requires that the non-Abelian  vector
currents be conserved, with the divergence of the non-Abelian
axial-vector currents   containing an anomaly.
In this renormalization scheme, ${\cal L}_{\rm WZ}$ has a form which
differs from that in the LR scheme  by a polynomial of the gauge
fields and their derivatives.
This can be expressed in terms of the corresponding effective
actions as
\cite{wzw}
\begin{eqnarray}
\Gamma^{\rm B}_{\rm WZ}(U,\ell,r) \;=\;
\Gamma^{\rm LR}_{\rm WZ}(U,\ell,r)
- \Gamma^{\rm LR}_{\rm WZ}(U=1,\ell,r)   \;.
\end{eqnarray}
The result  does not contain triple-gauge-boson vertices, so that the
analog of Eq.~(\ref{mwz}) vanishes and the low-energy
$\gamma^*\gamma Z_{\rm L}$ vertex  is determined by the leptons alone.
However, the calculation of the lepton-triangle vertex is different
in this scheme.
(This is the calculation usually illustrated in textbooks.)
Each lepton triangle has one axial-vector and
two vector vertices, and is evaluated by imposing vector-current
conservation.
The result is three times larger than Eq.~(\ref{ml}).
The total $\gamma^*\gamma Z_{\rm L}$ vertex at these energies is then
the same as
found previously, and agrees with Eq.~(\ref{pgg}).

In either form, the anomaly in the  divergences of
currents coupled to the electroweak gauge fields in the theory
vanishes, and hence does not disrupt gauge invariance,  because
contributions from the strong  and lepton sectors cancel,
as stated earlier.
This is due to the familiar fact that, for the fermion representation
considered, $\; {\rm Tr}(T_a\{T_b,T_c\}) = 0,\;$ where $T_a$, $T_b$,
and $T_c$ are the generators of the electroweak gauge group.

We have shown, in a model-independent way, that in a theory with
global anomalies  the triple-gauge-boson couplings due to the anomalies
are nonvanishing at low energies and consistent with the corresponding
couplings involving the Goldstone bosons, thereby satisfying
the equivalence theorem, in agreement with Kilgore's result.
This resolution implies the interesting possibility of
having anomaly-generated gauge-boson vertices in models of dynamical
electroweak symmetry breaking.
In studies on the effects of new, heavy physics on gauge-boson
self-interactions, it is usually assumed that there is no
contribution caused by anomalies.
However, there is no requirement that forbids the presence of global
anomalies due to new physics, provided that there are no anomalies in
the gauge currents in the theory.
Hence there may be TeV-scale theories of electroweak symmetry breaking
which do contain such anomalies.
These could then  generate contributions to gauge-boson vertices which
might have detectable consequences.

\bigskip

This work was supported in part by the U. S. National Science
Foundation.

\end{document}